\documentclass[12pt,longbibliography,eprint]{revtex4-1}
\usepackage{graphicx}
\usepackage{epstopdf}
\usepackage{hyperref}

\begin{document}
\preprint{V.M.}
\title{Features in Energy Accumulation in
  Double Layer on the surface of Graphene Material.
}

\author{Mikhail Evgenievich \surname{Kompan}}
\email{kompan@mail.ioffe.ru}
\author{Vladislav Gennadievich \surname{Malyshkin}} 
\email{malyshki@ton.ioffe.ru}
\author{Alexander Yurievich \surname{Maslov}}
\email{maslov@ton.ioffe.ru}
\affiliation{Ioffe Institute, Politekhnicheskaya 26, St Petersburg, 194021, Russia}
\author{Viktor Petrovich \surname{Kuznetsov}}
\affiliation{"NII GIRICOND" JS Co, 10 Kurchatova str., St.Petersburg, 194223, Russia}
\author{Viktor Aleksandrovich \surname{Krivchenko}}
\affiliation{Chemistry Department,
Moscow State University,
119991 Moscow Russia}

\date{November 3, 2015}

\begin{abstract}
\begin{verbatim}
$Id: quantlim.tex,v 1.16 2015/12/27 23:09:27 mal Exp $
\end{verbatim}
An application of quantum size carbon
structures--graphenes as electrodes of supercapacitors
is studied.
A fundamental limit of energy and power density arising
from quantum nature of objects
due to singularity in graphene  density of states
near Dirac point is determined
and technical solutions to partially offset
the negative factors are considered.
The maximum possible specific capacitance of nanostructured
electrode materials is determined.
\end{abstract}

\maketitle

\section{\label{intro}Introduction}

Devices and materials for energy storage are of paramount
importance to modern technology.
Requirements and the corresponding technical conditions
are so different that practically used technical solutions vary greatly.
Common approaches to a wide range of technical
solutions and examples can be found in the monographs\cite{barsukov2006new,yu2013electrochemical}.

Storage of electrical charge in double layer (EDCL)
is a fundamental principle of electrochemical capacitors, or supercapacitors.
Energy storage mechanism in these devices is different
from the mechanism in conventional capacitor;
However, since the total energy in both cases is proportional
to electrode area, the search of effective  material
for electrodes is an important problem in both cases.
Recent development led to numerous supercapacitor devices
with electrodes made of graphene-based
materials\cite{vivekchand2008graphene,mahmood2014graphene,liu2010graphene,kannappan2013graphene,wang2009supercapacitor}.
Nevertheless, none of these studies have reached the specific capacitance,
corresponding to known graphenes specific area 2630 $m^2/g$  \cite{mukhopadhyay2012graphite}
geometric limit.
We use the term "graphene--based materials" (GBM)
as a collective term for the materials studied in publications
on graphene application (for example \cite{singh2011graphene}),
as their properties are quite different\cite{huang2012overview} from ideal graphene,
but at the same time exhibit the main features of these objects.
In addition to possible technological factors,
there is exist capacity reduction\cite{luryi1988quantum,droscher2010quantum,nagashio2013estimation}
due to fundamental features of graphene.
Consequently, the effective specific capacity of the graphene material
is a function of accumulated charge and under certain conditions
can be substantially, more than an order of magnitude,
lower than that expected from specific surface estimation.
An effect of extra charge carriers, coming either from
doping the material or from charging the supercapacitor,
on specific capacity is exhibited.

\section{\label{theory}Capacitance}

In this paper we study the influence of the same factors
on energy characteristics of the graphene material.
Electrical energy accumulated on GBM
electrodes is equivalent\cite{kompan2015ultimate}
the energy of
to serially connected regular
(geometric, determined by electrodes surface) capacitance and
so called  quantum capacitance.
The differential capacitance $c(U)$ of
graphene sheet charged from an external source $U$  can be
expressed as
\begin{eqnarray}
  \frac{1}{c_T(U)}&=&\frac{dU}{dQ}=\frac{dU_g}{dQ}+\frac{dE}{dQ}
  \label{cq}
\end{eqnarray}
where $U_g$ is the electrostatic potential and $E$ is the
energy of electron (hole) in the conduction (valence)
band (here and below, it is assumed that the second
electrode of the capacitor has infinite capacitance).
The first term in Eq. (\ref{cq}) represents the typical
geometric capacitance ($\frac{dU_g}{dQ}=1/C_g$) and the second
term reflects the influence of quantum effects.
The total capacitance
can only be smaller the  "classical" $C_g$ one,
and the accumulated specific energy can be only lower.
The equivalence of GBM capacitance to serially connected
double layer capacitance and quantum capacitance
should have as a consequence not only the reduction in capacitance,
but also potential redistribution, as they are serially connected.

External voltage applied to the GBM capacitor
(measured at its electrodes) is a sum of $U_g$ (regular capacitor)
and $E$ (quantum capacitor).
The microscopic nature of these potentials is different.
$U_g$ voltage is related to geometric capacitance,
applied to a double charged layer at electrode surface--
electrolyte boundary.
$E$ - voltage "applied" to the quantum capacitance.
The origin  $E$ is charge redistribution within conduction band,
what lead to Fermi level shift and is equivalent to additional
$E$ potential applied. Given the density of states in the
conduction band this $E$  can be calculated
from the charge on the capacitor, as following.

\begin{figure}[t]
\includegraphics[width=7cm]{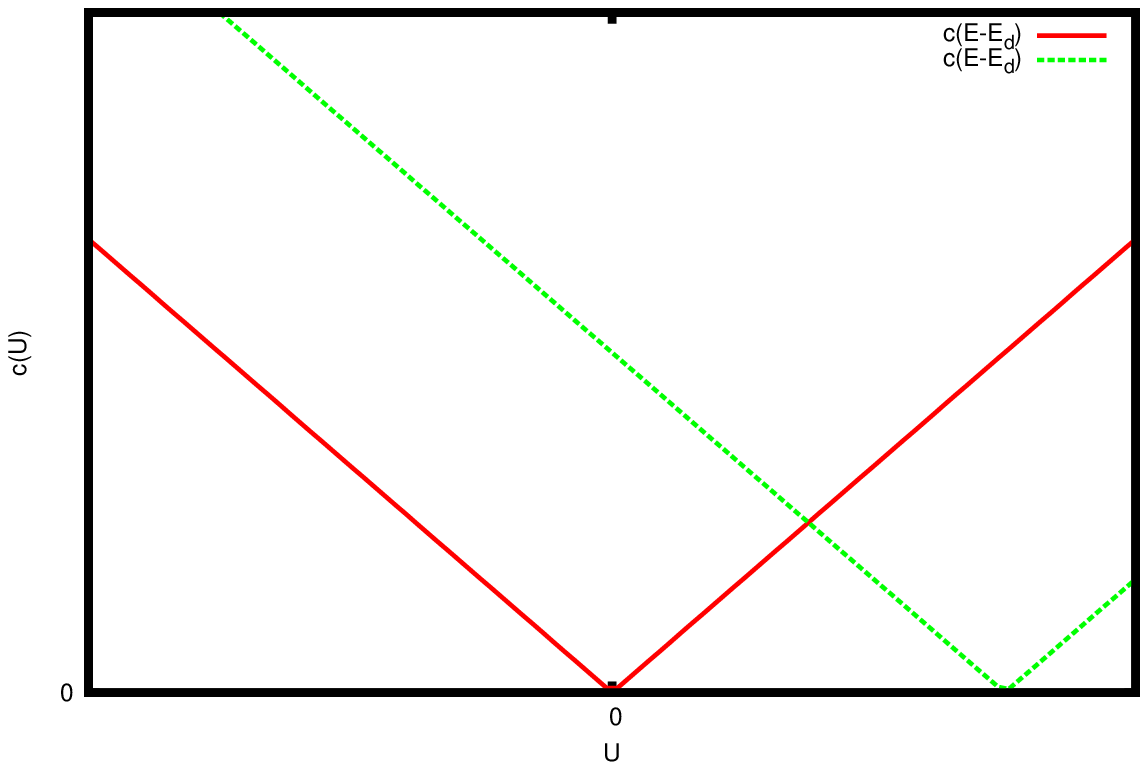}
\includegraphics[width=7cm]{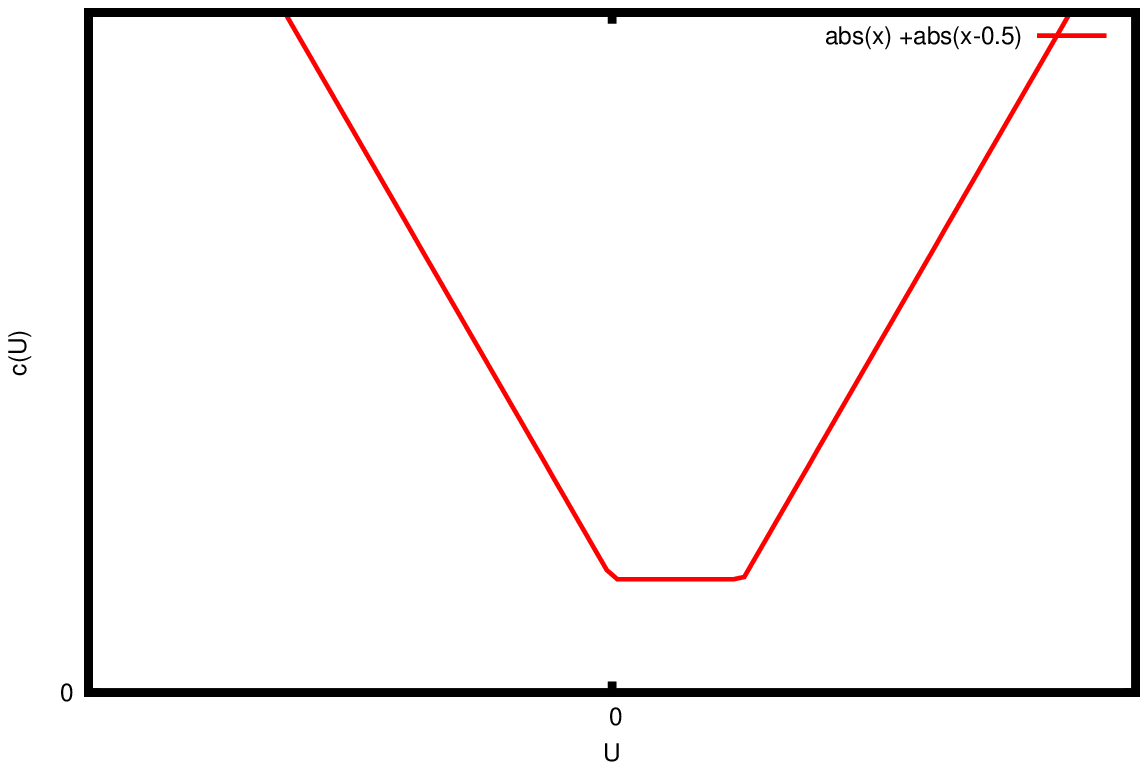}
\caption{\label{gdoped}
  Graphene $c(E)$  for different $E_d$ (left)
  and $c(E)$  for a system with two $E_d$ (right).
}
\end{figure}

Consider the quantum portion of differential capacity (\ref{cq})
\begin{eqnarray}
  \frac{dQ}{dE}&=&c(E) \label{cdiff}
\end{eqnarray}
In typical capacitors, the potential (energy per unit
charge) arises due to the electrostatic interaction of
accumulated electric charges. In the case of graphene,
the relationship between introduced charges and
required energy involves additional coupling that is
related to the finite electron density of states (DOS) in
graphene. The energy band structure of graphene has
a singularity at six equivalent points on the boundary
of the Brillouin zone. At these points, the boundary
states of the conduction band and valence band appear
as two cones sharing a common vertex\cite{neto2009electronic}. The DOS
of conduction electrons ($n$) and holes is zero at the
singular point and linearly increases with energy as
\begin{eqnarray}
  n(E) &=& \alpha E \label{nE}
\end{eqnarray}
Under not very high $E$ and Graphene quantum
linear dispersion relation
$c(E)=k\left|E\right|$ for undoped graphene,
and in homogeneously doped graphene
\begin{eqnarray}
  c(E)&=&k\left|E-E_d\right| \label{cdiffEF}
\end{eqnarray}
where the $E_d$ is doping level -- the difference
between Fermi level and singular point.
In case of inhomogeneously doped graphene the (\ref{cdiffEF})
should be averaged over $E_d$ and the result for two $E_d$
is presented in Fig. \ref{gdoped}.
For practical applications the total charge and energy accumulated
is of most interest. Integrating (\ref{cdiff}) from 0 to $E$
obtain
\begin{eqnarray}
  Q(E)&=&\int_0^{E} c(U)dU \label{QUq} \\
  W(E)&=&\int_0^{E} U c(U)dU \label{W}
\end{eqnarray}
The formulas (\ref{QUq}) and (\ref{W}) allows total charge $Q$
and energy $W$ to be calculated.
Consider the case when  charging (or discharging) current
is a constant $I$. Then $Q=It$ and total potential on
supercapacitor is a sum of regular double layer potential $U_g=Q/C_{dl}$
and Fermi level shift $E$, that (\ref{QUq})
on upper limit.
\begin{eqnarray}
  Q&=&It \label{It} \\
  Q&=&U_g/C_{dl} \label{QUg} \\
  U&=&E+U_g
\end{eqnarray}
The easiest way to calculate charging curve $U(t)$
is to calculate it in parametric form, using $E$ as a parameter,
calculating $U(E)$ and $t(E)$ from (\ref{QUq}), (\ref{QUg}) and (\ref{It}).
\begin{figure}[t]
  \includegraphics[width=7cm]{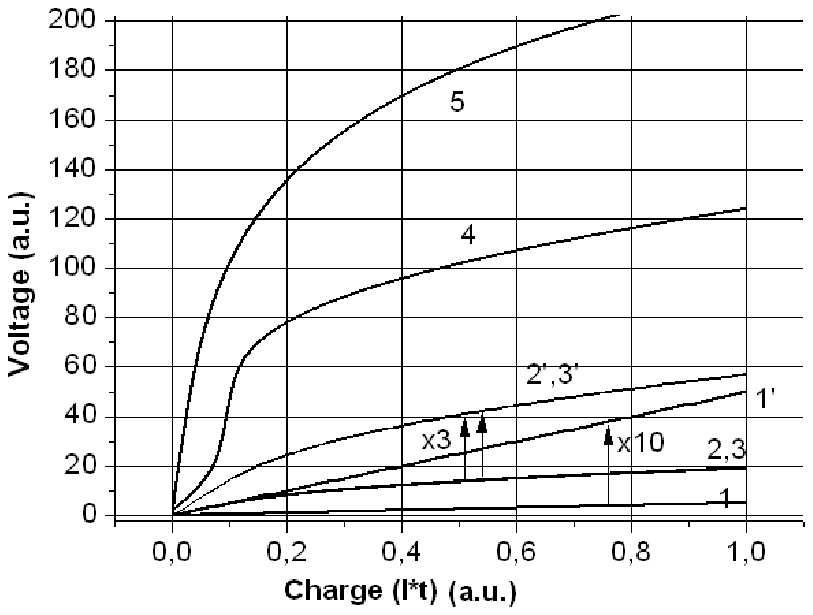}
  \includegraphics[width=7cm]{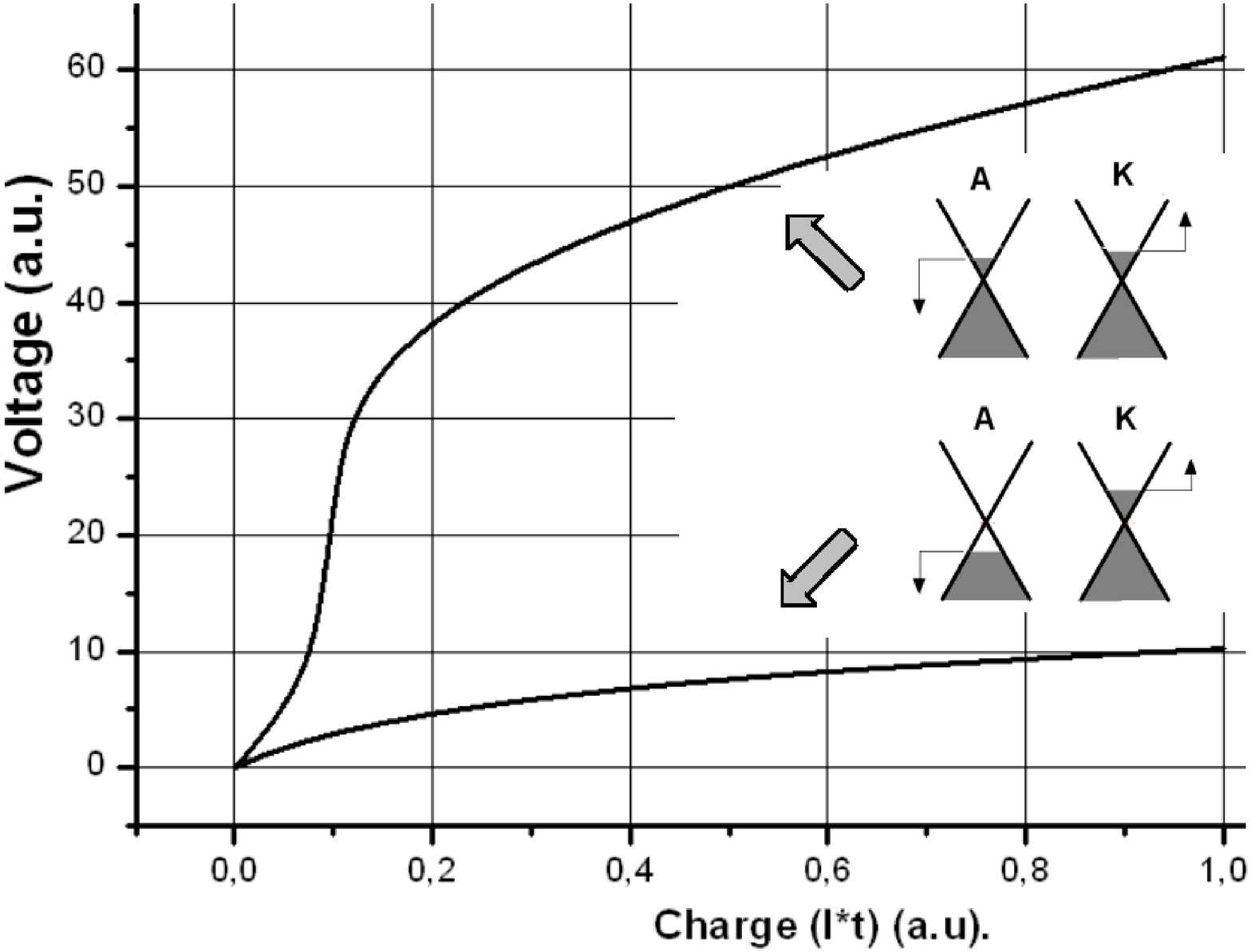}
\caption{\label{Ut}
  Left: $U_g(Q)$, $E(Q)$ and $U(Q)$ for supercapacitors with
  GBM electrodes.
  Right: $U(Q)$ for differently doped electrodes.
}
\end{figure}
The $U(Q)$, Fig. \ref{Ut}, because of an additional $E(Q)$ contribution to
$U_g(Q)$,  means capacity reduction
due to quantum effects. However, this quantum contribution $E(Q)$
depend on density of states at $U=0$
and can be greatly reduced by
doping supercapacitor electrodes as shown in Fig.\ref{Ut} (bottom).
Now additional charge would lead to a small
Fermi level shift, thus reducing $E(Q)$ contribution,
responsible for capacity reduction.
\begin{figure}
  \includegraphics[width=10cm]{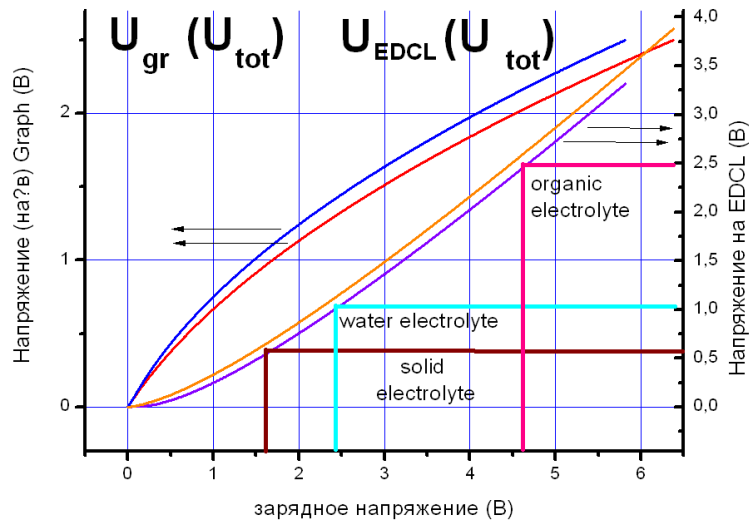}
\caption{\label{fig2}
  Potential redistribution on the supercapacitor.
}
\end{figure}
There is one more factor that limits accumulation
of charge and energy on EDLC - the  "Stability window"
of the electrolyte.
The voltage applied to double layer is limited
by electrolyte decomposition potential.
The limit is the maximal voltage not yet causing
electrolyte(liquid or solid) decomposition.
For solid electrolyte RbAg4I5 this value is about 0.55V;
for water-based electrolyte of about 1V, and for electrolytes
based on organic solvents limiting voltage typically is 2.5-3 V.
Important, that this limitation is applicable only
to double layer potential $U_g$.
There is no similar limitation mechanism applied to $E$.
Thus, the above discussion shows that the total voltage
that can be applied to the EDLC type under discussion,
is higher than the one determined from the "stability window"
on classical surfaces.

The calculations in Fig. \ref{fig2} were made under
the same conditions as in our paper\cite{kompan2015ultimate},
where we took into account graphene  density of states, what lead to dependence of the capacitance on the capacitor charge.
In Fig \ref{fig2}  the  $x$ is total voltage, and $y$ is the fraction,
applied to double layer (applied to electrolyte).
Fixing y value at electrolyte decomposing potential
then projecting to x one can obtain maximal potential,
which can be applied to the capacitor without electrolyte decomposition.
The graph shows that the total voltage can be up to 1.5--3
times greater than electrolyte decomposition potential.
The calculations show,
that in spite of the reduction in the effective capacity,
the effect can be partially compensated
by the maximal potential increase.

A family of curves in Figure \ref{fig2} reflects the redistribution
of potential applied to the GBM
(the lower axis, the potential  between the solution in volume and GBM)
between the voltage an the double layer (EDCL)
(right axis) and the rise of the Fermi energy in the GBM (left axis).
Two pairs of curves (a pair in each case)
on the figure correspond to the two types GBM doping\cite{kompan2015ultimate}.

\section{\label{power}Power}

Another extremely important parameter of energy storage device
is the maximal output power. The parameter(besides voltage)
determining the power is internal resistance.
The internal resistance is often limited by
the conductivity of electrodes material and
device technology. Electrolyte   conductivity,
while important, is more difficult to optimize.
A distinctive feature of EDLC on GBM should be the dependence
of material conductivity on accumulated charge.
The reason for this dependence should be the same
density of state dependence on accumulated charge.
In the simplest case the potential drop on internal resistance
would be:
\begin{eqnarray}
  U_R&=&I/\sigma(E) \label{UR} \\
  U&=&E+U_g+U_R \label{UU}
\end{eqnarray}
where the conductivity $\sigma(E)$ depend on
carrier density through the $E$ potential, same as used for
capacity calculations.
In the Eq. (\ref{UU})
the internal resistance term is added to total potential.
In the simplest case the $\sigma(E)$ dependency can be estimated
from simple considerations. Standard graphene
dispersion near band edge $E=\hbar V_F |k|$
have zero effective mass and almost infinite mobility.
However, this is the case only near the band edge.
In the band is partially occupied, to the level of $E$,
then the effective mass $m^*$, affecting the mobility, can be estimated as
\begin{eqnarray}
  m^*&=&\frac{e E}{V_F}
  \label{effmass} \\
  \mu&=&\beta \frac{1}{E} \label{mu}
\end{eqnarray}
and growth with filling the band. Thus the dependence
of conductivity on $E$ has two factors:
growing with $E$ the concentration of carriers (\ref{nE})
as $\left|E\right|$
and declining as $1/\left|E\right|$
mobility due to effective mass (\ref{effmass}) increase.
For power measurement
then the $\sigma(E)$ take a constant value.
In the case of graphene $\sigma(E)$
exhibit no singularity near $E=0$,
because of singularities cancellation in numerator and denominator.
This effect is different from typically studied
in graphene Dirac point conductivity, that
consider small current experiments.
Here on large current, most interesting setup
for supercapacitors, the effective mass expression(\ref{effmass})
is different and give as a result finite conductivity.

\section{\label{discussion} Discussion}
In this work limit capacitance, maximal potential and power
of supercapacitors with electrodes made from GBM are estimated.
It was shown that quantum capacity limit
total capacity, but the effect can be partially offset by
1) Doping graphene electrodes to increase carriers concentration
and 2) Due to potential redistribution between double layer
and band shift total voltage, that can be applied to
the device become higher than in classical case, thus
allows partial offset of capacity reduction.

\bibliography{echem}

\end{document}